\documentstyle[preprint,tighten,version2,aps]{revtex}
\input{epsf.tex}
{\makeatletter%
\gdef\labeleqs#1{{%
\edef\@currentlabel{%
\ifappendixon\appletter\fi
\ifsecnumbers\ifnum\c@secnum>0 
\arabic{secnum}.\fi\fi\arabic{equation}}%
\label{#1}%
}}%
} 
\begin{document}
\draft
\preprint{IFUP-TH 13/96}
\begin{title}
Condensation of vortices in the X-Y model in 3d: a disorder parameter.
\end{title}
\author{G. Di Cecio\thanks{Presently at: Dpt. of Physics, Louisiana State
University, Baton Rouge.}, A. Di Giacomo, G. Paffuti and 
M. Trigiante\thanks{Presently at: SISSA, Trieste }}
\begin{instit}
Dipartimento di Fisica dell'Universit\`a and 
I.N.F.N., Pisa, Italy 
\end{instit}
\begin{abstract}
A disorder parameter is constructed which signals the condensation of vortices.
The construction is tested by numerical simulations.
\end{abstract}
\pacs{PACS numbers: 75.40.Cx, 05.70.Jk, 64.60.Cn}
\section{Introduction}
The $XY$ model in 3 dimensions is an interesting system in statistical 
mechanics. Physically it describes the critical behaviour of superfluid
${\rm He}_4$\cite{1}. 
From the theoretical point of view it provides a relatively
simple example of an order-disorder phase transition in which the condensation of
solitons plays an essential role[2-5]. It is fairly well established that
the transition is continuous, and the basic critical indices are known with good
precision\cite{6,7,8}.

A Ginzburg-Landau kind of description of the system has been given, describing
phenomenologically the condensation of vortices in the high temperature 
phase\cite{9}.

In this paper we shall study the system from a slightly different point of
view. We shall look at it as the lattice formulation of a 2 + 1 dimensional
quantum field theory. We shall then show that the (2 dimensional) ground state
of the theory spontaneously breaks the conservation of the number of vortices
at high temperature; vortices condense in the vacuum in the same way as Cooper
pairs do in a superconductor.
In that phase
the number of vortices is not defined in the vacuum, which is a superposition
of states with different vorticity. We shall  directly exhibit this
condensation by showing that the vacuum expectation value ({\it vev}) of an
operator which changes the number of vortices is non zero.

An analogous technique has been used to detect monopole condensation in
the ground state of $U(1)$ gauge theory\cite{10} and in $SU(2)$ gauge theory, as
a mechanism for confinement of colour\cite{11}.

In sect. 2 we shall briefly introduce the model and fix the notation.
We shall also define a creation operator for a vortex.

In sect. 3 we shall present and discuss the results of numerical simulations
where the  {\it vev} of the above operator is measured. The
result will be a direct detection of the spontaneous breaking of the
conservation of the number of vortices in the disordered phase and
a direct evidence of vortex condensation.
We shall also obtain an alternative determination of known critical indices, 
which
agrees with existing results, and measure the critical index related to our
disorder parameter.

\section{The creation operator of vortices.}
The lattice action of the system is
\begin{equation}
S = \beta \sum_{\mu,i}\left(1-\cos(\Delta _\mu \theta(i) )\right)\end{equation}
The field variable is the angle $\theta$ in  the site $i$: $\mu$ runs from 0 to
2, 0 being the time axis. 

In the limit of zero lattice spacing ($a\to 0$)
\begin{equation}
S \simeq \frac{\beta}{2}\sum_{\mu,i} \left(\Delta_\mu \theta\right)^2 a^2 +
{\cal O}(a^4)\end{equation}
At high $\beta$'s the system describes a massless free scalar particle. The
couplings of higher dimension appearing in the series expansion of the cosine
become important at lower $\beta$. At $\beta_c\simeq 0.454$ the system
undergoes a phase transition, which is known to be of second order~\cite{6}.

The system has soliton configurations with the geometry of vortices: numerical
simulations show that vortices play an important role in the phase 
transition\cite{3,4,5}.

A vector field 
\begin{equation} A_\mu = \partial_\mu\theta\label{eq:defa}
\end{equation}
 can be defined, and  a current $j_\mu = \varepsilon_{\mu\alpha\beta} 
\partial_\alpha A_\beta$
associated with it,
which is trivially conserved in smooth configurations:
\begin{equation}
\partial^\mu j_\mu = \partial^\mu \varepsilon_{\mu\alpha\beta} \partial^\alpha
A^\beta = 0\label{eq:conscurr}\end{equation}
because of the antisymmetry of the Ricci tensor. Eq(\ref{eq:conscurr}) is the
analog of the Bianchi identity in QED.

The constant of the motion is
\begin{equation}
\Phi = \int{\rm d}^2x\, j^0(\vec x,x^0) = \int{\rm
d}^2x\,\left(\vec\nabla\wedge\vec A\,\right)\end{equation}
On the other hand from eq.(\ref{eq:defa}) 
\begin{equation}
\Phi = \oint_{\cal C}\vec A\cdot{\rm d}\vec x =
n\cdot2\pi\label{eq:number}\end{equation} 
with ${\cal C}$ any closed path.

A vortex is defined as a
configuration for which $n$ is non zero. 
\begin{equation}
\theta(\vec x - \vec y) = {\rm arctg}\frac{(x-y)_2}{(x-y)_1}\end{equation}
is an example of such a configuration with $n=1$. It has a singularity at $\vec x
= \vec y$.

Eq.'s (\ref{eq:defa}-\ref{eq:number}) tell that the number of vortices is a
constant of motion, and define a $U(1)$ symmetry.

We shall show that, for $\beta > \beta_c$ this symmetry is realized \`a la
Wigner: for $\beta < \beta_c$ a spontaneous breaking of it occurs. 

To do that
we will define an operator which creates a vortex\cite{10}. In the continuum
version of the theory $\partial_0\theta$ is the conjugate momentum to the field
variable
$\theta$. To add a vortex in the site $\vec y$, $\bar\theta(\vec x-\vec y)$ to
any field configuration the analog of the simple translation
\[ {\rm e}^{{\rm i}p a}| x\rangle=|x+a\rangle\]
can be used. The translation operator
\begin{equation}
\mu(\vec x,t) = \exp\left[{\rm i}\int{\rm d}^3\vec y\partial_0\theta(\vec y,t)
\bar\theta(\vec x-\vec y)\right]\label{eq:8}\end{equation}
indeed, when applied to the state $|\theta(\vec x,t)\rangle$ gives
\begin{equation}
\mu(\vec x,t)|\theta(\vec y,t)\rangle = |\theta(\vec y,t) + \bar\theta(\vec
x - \vec y) \rangle\end{equation}
The operator number of vortices
\begin{equation}
V(t) = \int{\rm d}^2x\,j^0(\vec x,t)\end{equation}
has commutation rule with $\mu$
\begin{equation}
\left[V(t),\mu(\vec x,t)\right] = 
\int{\rm d}^2z\,\vec\nabla\wedge\vec A(\vec
z,t)\,\mu(\vec x,t) =
2\pi\mu(\vec x,t)\end{equation}
where $\vec A = \vec\nabla\bar\theta$.

If the $U(1)$ symmetry which counts the vortices is realized \`a la Wigner,
then the ground state must have a definite number of vortices and 
$\langle \mu\rangle = 0$. Since $\mu$ changes the number of vortices
a {\it vev} $\langle \mu\rangle \neq 0$ signals their condensation.

The naive translation of eq.(\ref{eq:8}) on the lattice is\cite{10}
\begin{equation}
\mu(\vec n,n_0) = \exp\left[-\beta\sum_{\vec
n'}\sin(\Delta_0\theta(\vec n',n_0))\bar\theta(\vec n'-\vec n,n_0)\right]
\label{eq:8a}\end{equation} 
where the site $\vec n'$ runs on the sites of a 2 dimensional slice of the
lattice at constant
$n_0$, except the position of the vortex.

Instead of Eq.(\ref{eq:8a}) we shall
use a compactified version of it. By this we mean
\begin{equation}
\mu = \exp\left[-\beta\sum_{\vec n'}\left(
\cos(\Delta_0\theta +\bar\theta)- \cos(\Delta_0\theta)\right)\right]
\label{eq:8b}\end{equation} 
which coincides with eq.(\ref{eq:8b}) at first order in $\bar\theta$.
With the definition (\ref{eq:8b}) $\langle\mu\rangle$  
is independent of the choice of the gauge for $A_\mu$, i.e. of the zero for the
angle $\theta$: a redefinition of it is reabsorbed in the integration variable.

Two remarks are in order
\begin{itemize}
\item[1)] in principle one expects $\langle\mu\rangle = 0$ in the ordered
phase $\beta > \beta_c$. In fact $\langle\mu\rangle$ is an analytic function
of $\beta$ if the number of degrees of freedom is finite, and therefore, if it
were zero for $\beta > \beta_c$ it would be identically zero for all $\beta$'s.
Only in the limit of infinite volume (thermodynamic limit) singularities
develop, and $\langle\mu\rangle$ can be identically zero above $\beta_c$,
without being zero everywhere.
Indeed
a simple computation at large $\beta$'s, where the integral is approximately
gaussian, gives 
\begin{equation}
\langle\mu\rangle \sim \exp\left[-\beta\left( c_1 V^{1/3}
+ c_2 + {\cal O}(\frac{1}{\beta}\right)
\right]\label{eq:gaussian}
\end{equation}
\item[2)]
Due to periodic boundary conditions the total number of vortices must be zero as
easily seen from eq.(\ref{eq:number}) with the boundary $C$ at spatial infinity.
In principle then the correct way to determine $\langle \mu\rangle$ is 
to observe a vortex antivortex pair correlation
\begin{equation}
\langle \mu(\vec x) \,\mu (\vec 0)\rangle \mathop\sim_{x\to \infty} \langle \mu
\rangle^2
\end{equation}
By cluster property and $C$
invariance it tends exponentially to $\langle\mu\rangle^2$ at large distances.

In practice, since in the thermodynamical limit physics becomes independent of
the boundary conditions and decorrelated from them at distances larger
than the correlation length, what happens is that also a single vortex can be
created, and a direct measurement of $\langle\mu\rangle$ can be done.
$\langle\mu\rangle$ determined in this way is indeed independent of the
boundary conditions within errors. When a single vortex is put on the lattice
by our operator
a dislocation appears at the boundary with opposite vorticity
to correct for the boundary conditions, and, 
if the lattice is big compared to the correlation length,
what is obtained is 
the same $\langle\mu\rangle^2$  as with a vortex antivortex pair at large
relative distance.
\end{itemize}

\section{Numerical simulation and results.}
Simulations have been done on a $20^2\times40$ and $30^2\times 60$ lattices to
measure the $\langle \bar \mu \, \mu\rangle$ correlation. Single vortices
$\langle\mu\rangle$ have also been measured 
on the same lattices and
on  $20^3$, $30^3$ and $40^3$
lattices. Instead of directly determining $\langle\mu\rangle$ as a function of
$\beta$ it is more convenient to measure\cite{10}
\begin{equation}
\rho = \frac{1}{2}\frac{{\rm d}}{{\rm
d}\beta}\ln\langle\mu\rangle^2\end{equation} 
and to reconstruct $\langle\mu\rangle^2$
as
\begin{equation}
\langle\mu\rangle^2 = \exp\left[2\int_0^\beta\,\rho(\bar\beta)\,{\rm
d}\bar\beta\right]\label{eq:valmu}
\end{equation}
The observed shape of $\rho$ is shown in fig.1. The sharp negative peak at
$\beta_c$ indicates a steep drop of $\langle\mu\rangle$ toward zero at $\beta_c$.
At low
$\beta$'s
$\rho$ 
is compatible with 0
and independent of $V$ (fig.2). This means by eq.(\ref{eq:valmu}) that
$\lim_{V\to\infty}\langle\mu\rangle \simeq 1$ for $\beta < \beta_c$, and thus
that vortices do condense in the vacuum. At $\beta > \beta_c$ the behaviour of
eq.(\ref{eq:gaussian}) can be checked by numerical calculation of the gaussian
integral. For a cubic lattice with a monopole in the center the result is 
\begin{equation}
\rho = -11.332\cdot L + 72.669 + {\cal O}(1/\beta)
\label{eq:pertrho}
\end{equation}
The comparison with numerical simulations is shown in fig.1.

Eq.(\ref{eq:pertrho}) means that $\langle\mu\rangle^2$ is exactly zero in the
limit $V\to \infty$. Eq.(\ref{eq:pertrho}) agrees with the observed values of
$\rho$ at large $\beta$'s (fig.1).

Near $\beta_c$ where the correlation 
length goes large, a finite size analysis can be done to explore the limit of
infinite volume and determine the critical index $\delta$ of $\mu$, $\beta_c$,
and the critical index $\nu$ of the correlation length.

As $(\beta-\beta_c)\to 0^-$ and $V \to \infty$
\begin{equation}
\langle\mu\rangle\sim (\beta_c-\beta)^\delta\end{equation}
At finite $L$,
\begin{equation}
\langle\mu\rangle = \langle\mu\rangle\left(\frac{\xi}{L},\frac{a}{\xi}\right)
\end{equation}
As $\beta\to\beta_c$
 $\xi$ diverges, $\xi\sim (\beta_c-\beta)^{-\nu}$
and the dependence on $a/\xi$ can be
neglected
\begin{equation}
\langle\mu\rangle \simeq f\left( L^{1/\nu}(\beta_c-\beta)\right)\end{equation}
or
\begin{equation}
\rho \simeq L^{1/\nu}f\left( L^{1/\nu}(\beta_c-\beta)\right)
\end{equation}
$\rho/L^{1/\nu}$ in the scaling region must be a universal function of
$L^{1/\nu}(\beta_c-\beta)$. The quality of this scaling is shown in fig.3. The
critical index $\nu$ and the critical temperature $\beta_c$ are determined by
enforcing this scaling behaviour. 

Fig.4 shows the data for a vortex antivortex
pair at large distance, compared to the single vortex data. 
In the condensed phase the single vortex behaves
in fact as a pair at large distance, confirming that there is no problem in the
thermodynamical limit. 
At $\beta>\beta_c$ the two quantities are different and at $\beta$ sufficiently
large are well described by eq.(\ref{eq:gaussian}) with the different appropriate
values of
$c_1$ and $c_2$, as shown in the figure.

A more precise finite size scaling analysis assuming 
a specific form for $f$ for $\beta < \beta_c$
\begin{equation}
f(L^{1/\nu}(\beta_c-\beta)) = \frac{\displaystyle \delta}{
\displaystyle \left[\left(L^{1/\nu}(\beta_c-\beta) + v_1\right)^2 +
v_0^2\right]^{1/2}}
\label{eq:formoff}
\end{equation}
gives a precise determination of the critical
indices and of $\beta_c$
\begin{eqnarray}
\delta &=& 0.740 \pm 0.029\\
\beta_c &=& 0.4538 \pm 0.0003 \\
\nu  &=& 0.669 \pm 0.065\end{eqnarray}
with $\chi^2/dof = 1.07$.
For comparison the determination of ref.[8] is $\nu=.670(7)$ and
$\beta_c = .45419(2)$.

The behaviour of the correlator $\langle\mu(x)\bar\mu(0)\rangle$ as a function
of the distance gives an estimate of the correlation length $\xi_M(\beta)$
\begin{equation}
\langle\mu(x)\mu(0)\rangle -\langle\mu\rangle^2
\simeq\exp\left[-x/\xi_M(\beta)\right]
\end{equation}
$\xi_M$ is of the order of a few lattice spacings (see fig.5).
$\xi_M(\beta)$ should scale as $(\beta_c-\beta)^{-\nu}$ as 
$(\beta-\beta_c)\to 0^-$.
With our relatively large statistical errors we were not able to check this
behaviour.

\section{Concluding remarks.}
We have constructed a disorder parameter for the $\lambda$ transition in the
$3d$
$XY$ model which is the {\it vev} of an operator with non trivial vortex number.
It is different from zero at high temperatures ($\beta < \beta_c$)
in the disordered phase, thus demonstrating condensation of
vortices. 
In the ordered phase it tends to zero
exponentially with the linear size of the lattice. 

$\beta_c$ and the critical index $\nu$ of the correlation length can be
determined with good precision by a finite size scaling analysis of our
disorder parameter $\langle\mu\rangle$. As $(\beta-\beta_c)\to 0^-$
$\mu\sim(\beta_c-\beta)^\delta$ with $\delta = .74 \pm .03$.

A similar construction also works for other models undergoing an order disorder
transition produced by condensation of solitons. An example is the condensation
of monopoles in gauge theories\cite{10,11}.

\vfill\eject
{\centerline{\Large Figure Captions}}
\begin{itemize}
\item[Fig.1] $\rho$ as a function of $\beta$. The dashed lines at high values
of $\beta$ are the perturbative estimates.
\item[Fig.2] $\rho$ as a function of $1/L$ ($L$ = size of the lattice) in the
condensed phase. $\rho$ converges to a value consistent with zero. The abscissae
for different betas are slightly displaced for graphycal reasons.
\item[Fig.3] Quality of the finite size scaling analysis. The figure
corresponds to optimal values of $\beta_c$ and $\nu$.
\item[Fig.4] Comparison of $\rho$ for single vortex to $\rho$ for
vortex - antivortex pair at large distance. The two curves coincide within
errors in the condensed phase. They differ at high $\beta$ and agree with the
perturbative determination (dashed lines on the right side).
\item[Fig.5] $\rho$ as a function of distance at two values of $\beta$. The
estimated correlation length is few lattice spacings.
\end{itemize}
\vskip0.3in
\par\noindent
{\centerline{\epsfbox{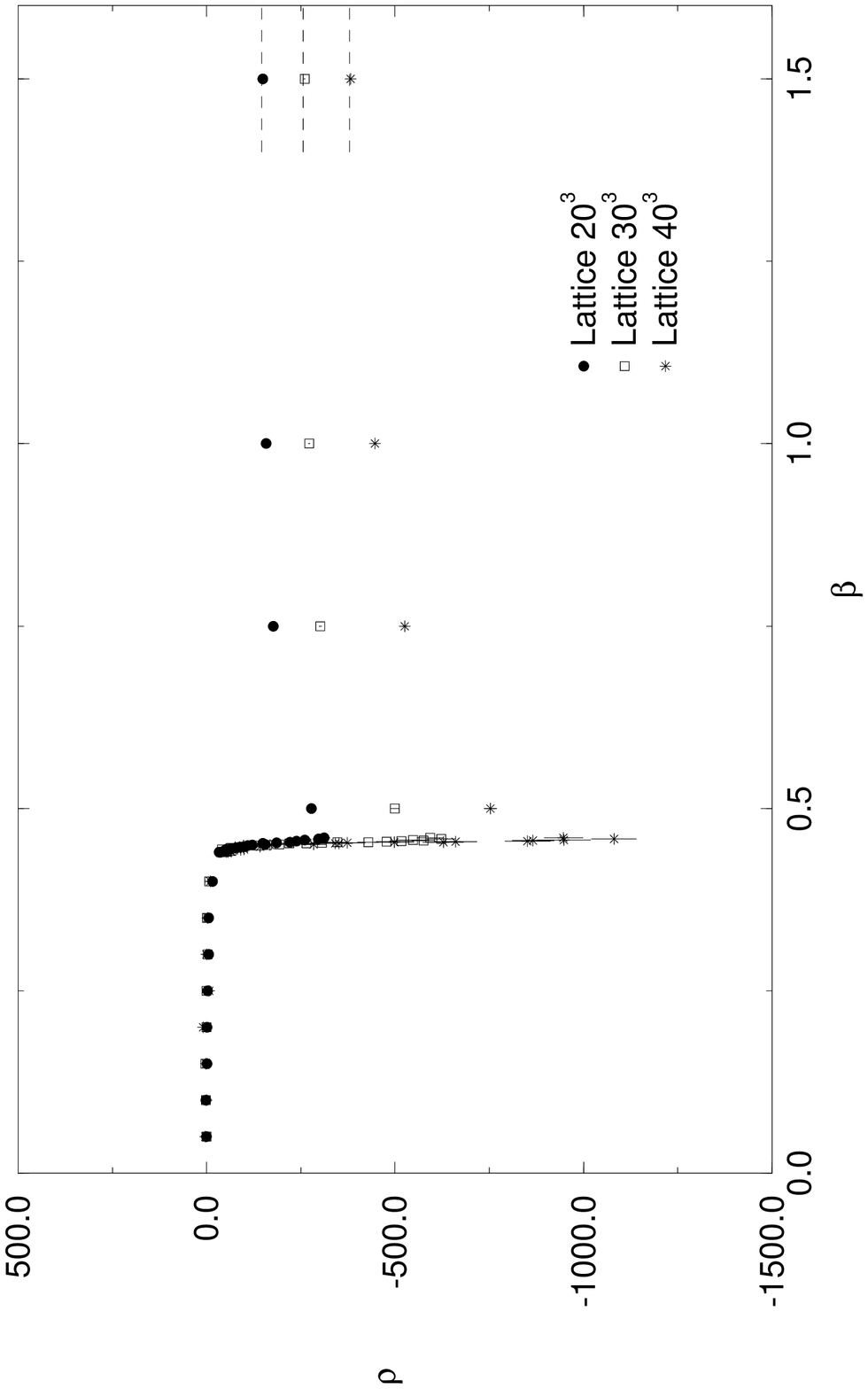}}}
\vfill\eject
\vskip0.3in
\par\noindent
{\centerline{\epsfbox{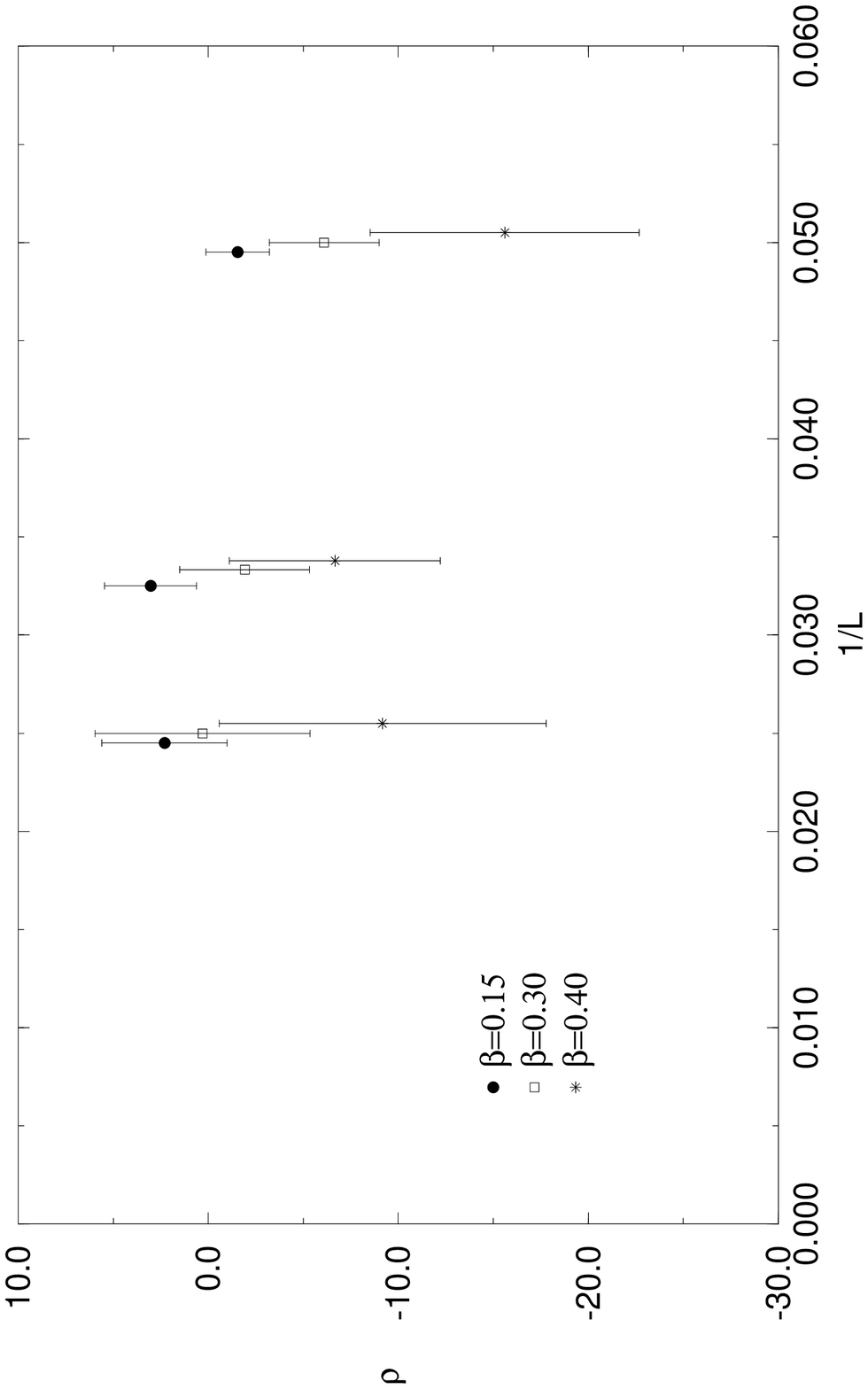}}}
\vfill\eject
\vskip0.3in
\par\noindent
{\centerline{\epsfbox{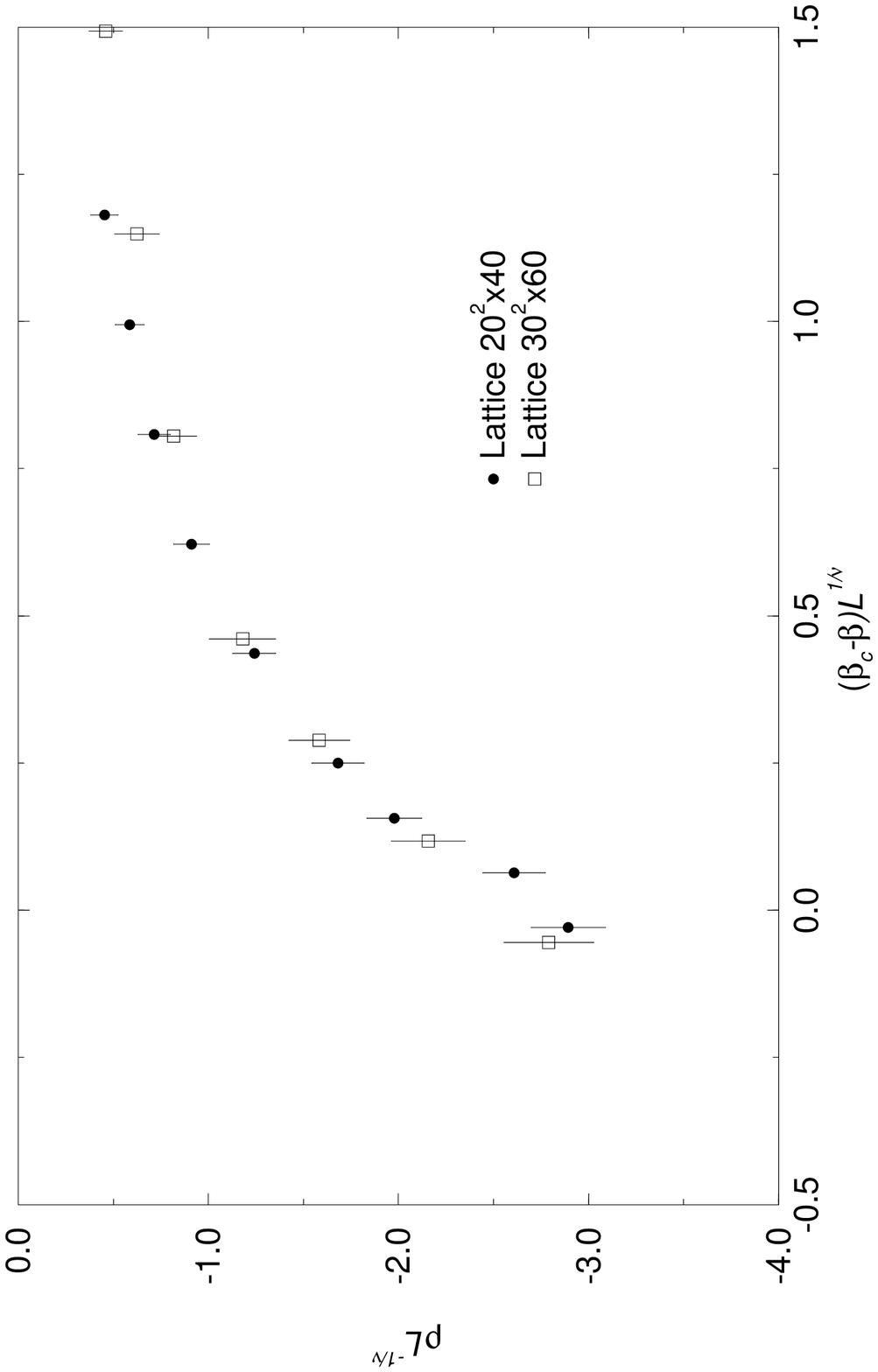}}}
\vfill\eject
\vskip0.3in
\par\noindent
{\centerline{\epsfbox{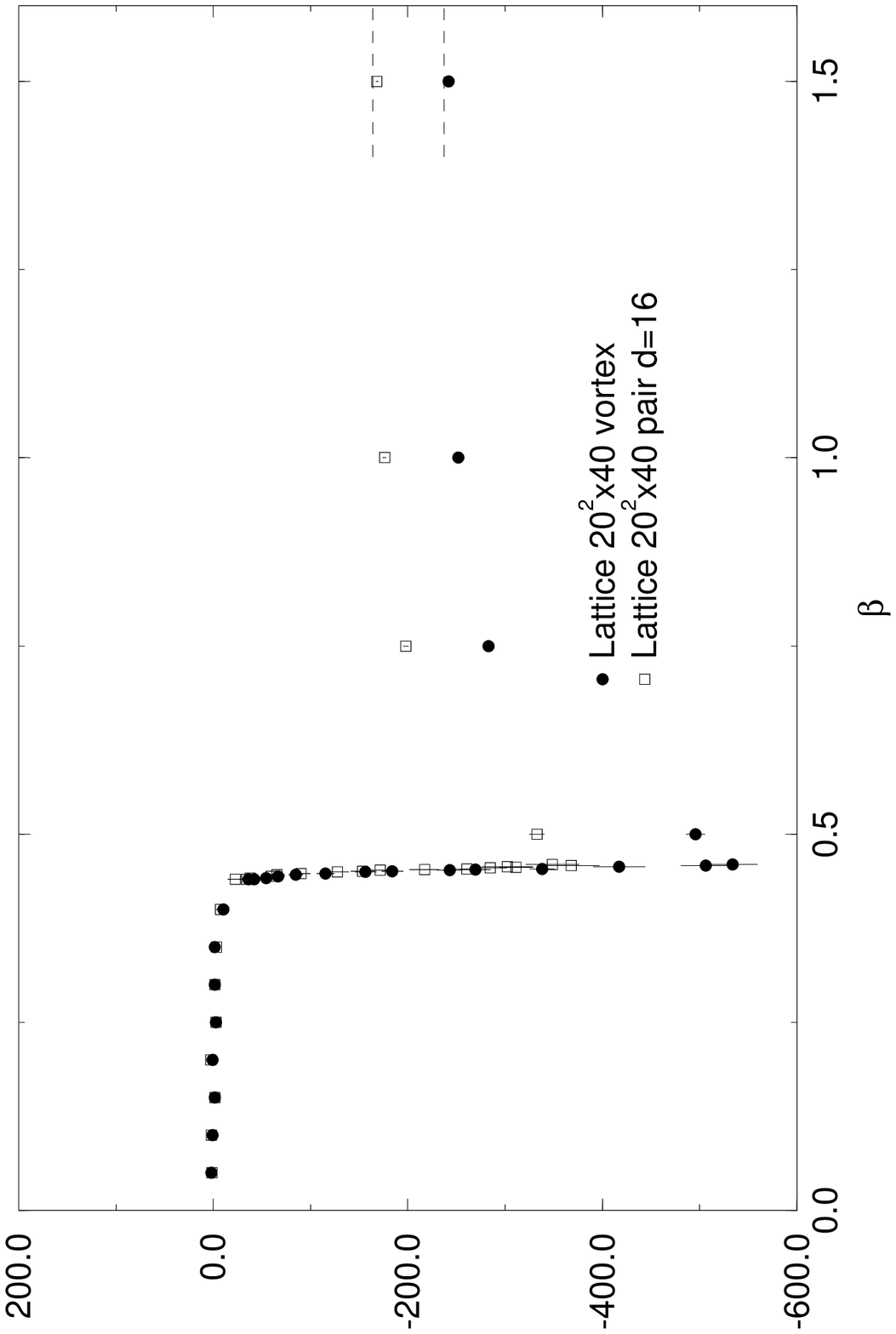}}}
\vfill\eject
\vskip0.3in
\par\noindent
{\centerline{\epsfbox{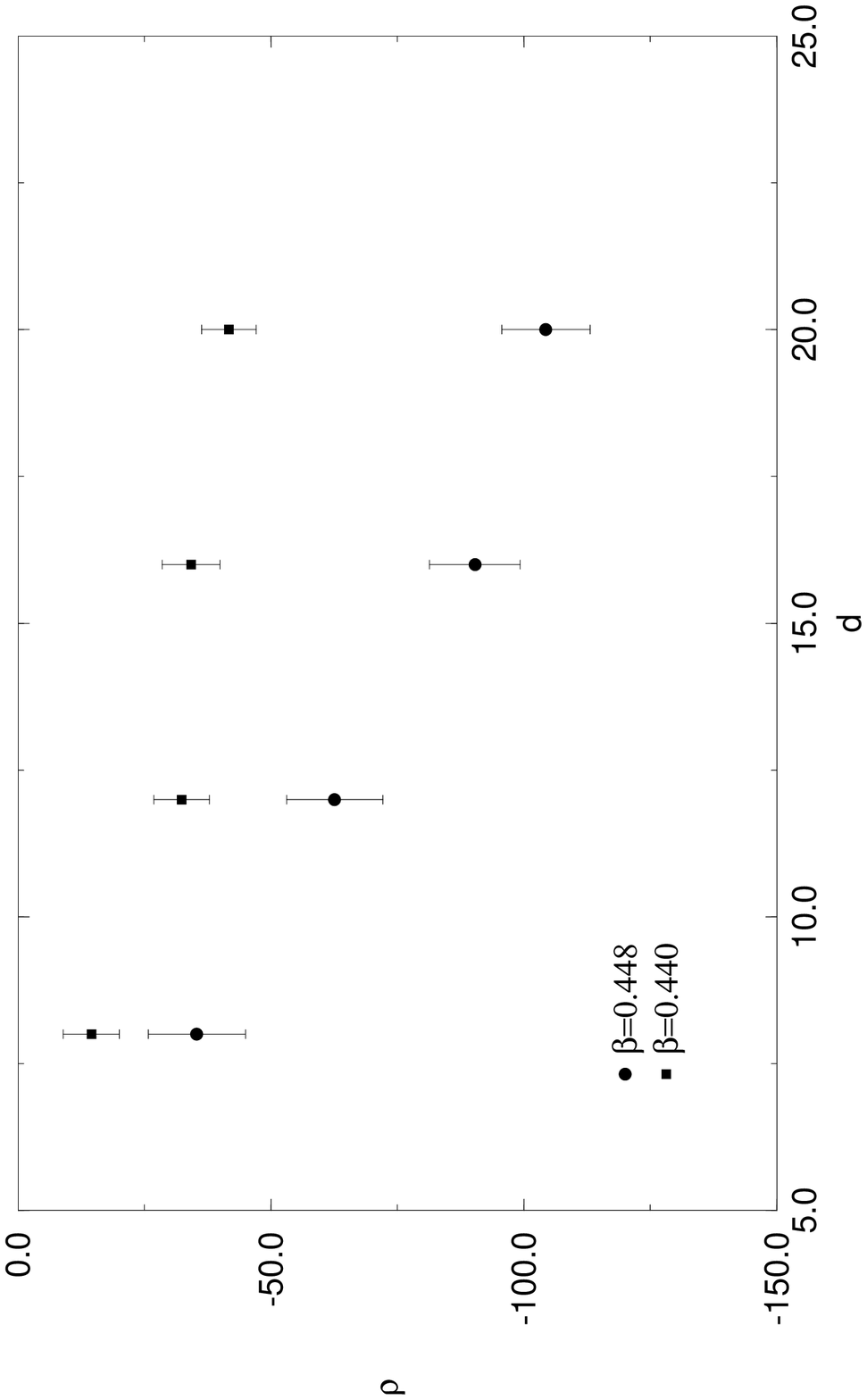}}}
\end{document}